\documentclass[a4paper,fleqn,usenatbib]{mnras}

\usepackage[T1]{fontenc}
\usepackage{ae,aecompl}
\usepackage{graphicx}   % Including figure files
\usepackage{amsmath}    % Advanced maths commands
\usepackage{amssymb}    % Extra maths symbols
\usepackage{xcolor} 
\usepackage{graphicx,bm,times,subfigure}   % for markup

\newcommand{\source}{\hbox{NGC\,6109}}
\newcommand{\chandra}{\textit{Chandra}}
\newcommand{\rosat}{\textit{ROSAT}}

\title{Extreme jet bending on kiloparsec scales : the `doughnut' in NGC 6109}

\author[Josie Rawes, Mark Birkinshaw, Diana M Worrall]
 {Josie Rawes, Mark Birkinshaw, Diana M Worrall
\\
HH Wills Physics Laboratory, University of Bristol, Tyndall
Avenue, Bristol BS8~1TL \\
}

\begin{document}

\label{firstpage}

\maketitle

\begin{abstract}

\noindent We present new radio observations of the $z$ = 0.029 radio galaxy NGC 6109, a member of the 3CRR sample. We find the radio morphology of the counter-jet to be highly distorted, showing a unique `doughnut' structure $\sim$6 kpc in diameter. The doughnut is overpressured compared with the surrounding atmosphere as measured with \chandra. We investigate the polarisation properties of the source and find evidence for an interaction between the doughnut and the external environment. This may cause the extreme jet bend. Alternatively, while providing no explanation for the rotation-measure and magnetic field structure seen in the doughnut, a ballistic precession model may be feasible if the ballistic flow persists for a distance much less than the full extent of the 100 kpc-scale jet. A light jet being deflected by gas flows and winds just outside the transition between the galaxy and cluster atmospheres appears to be a more plausible interpretation.  

\end{abstract}


\section{Introduction}

Curved radio structures and bent jet trajectories on kpc scales are frequently observed extending away from AGN. In some cases these bends appear helical in nature, with sinusoidal trajectories and oscillating ridge lines (e.g. NGC 315, \citealp{worrall07}; 3C 273, \citealp{romero00}). Theoretical models for such oscillating, bent structures include helical modes in hydrodynamic (\citealp{hardee87}; \citealp{birkinshaw91}) or magnetised \citep{konigl85} jets. The collision of a jet with a dense or magnetised medium is also capable of producing strong deflections, as seen in NGC 7385 \citep{rawes15}. Alternatively, in both Galactic and extragalactic jets, precession may be a viable mechanism for the production of observed helical signatures (e.g. \citealp{linfield81}; \citealp{gower82}). Most mechanisms suggested for causing jet precession involve creating a misalignment between the accretion disk and the black hole, either through a supermassive black hole binary system (e.g. 4C 73.18, \citealp{roos1993}; OJ 287, \citealp{lehto96}; Cygnus A, \citealp{canalizo03}), or an offset in the angular momentum direction between the accretion disk and the host galaxy. 

The 3CRR radio sample \citep{laing83} contains 35 low redshift (z $<$ 0.1) radio galaxies. How these galaxies interact with the external environment has been studied in depth for the majority of sources with the aim (among others) of investigating how radio sources are modified by the effect of the intergalactic medium (IGM) and how the IGM is modified by the interaction. A few sources, however, lacked adequate archival radio data for good radio/X-ray comparison. Recent Karl G. Jansky Very Large Array (VLA) observations of two head-tailed galaxies within this sample reveal strong distortions in the radio morphology, and magnetic features in the jet bends. The results for NGC 7385 are given in Rawes et al. (in prep). NGC 6109 is discussed here. 

NGC 6109 is an FRI-type head-tail radio galaxy located in a poor cluster with redshift z = 0.029 \citep{wegner}. It was mapped with the Westerbork Synthesis Radio Telescope (WSRT) and the extended radio structure was reported by \cite{ekers78}. The radio tail extends for a projected distance of 250 kpc to the NW, and it was noted by \cite{ekers78} that there may be a small component SE of the core of NGC 6109. A high resolution map was published in 1985 in a sample of 57 narrow angle tailed sources \citep{odea85}. The map showed a circular component SE of the core, with no obvious counter-jet. \cite{odea85} suggested that this structure might be produced if the beams were ejected parallel and antiparallel to the direction of motion of the galaxy, in a similar way to NGC 7385 \citep{schilizzi75}. They proposed that the counter-jet ejected in the direction of motion would interact with the ICM in a manner more typical of double radio sources and might form an edge-brightened lobe. 

New VLA observations presented in this paper reveal that the counter-jet exhibits a highly unusual loop. Such a morphology has not been observed for any other low power radio galaxy. We discuss whether jet precession or gas-dynamical jet bending could produce this unique structure. We use our polarisation data to probe the magnetic field structure of this radio component and we investigate the local X-ray environment with \chandra\ observations.

Throughout this paper we adopt the cosmological parameters H$_0$= 70 km s$^{-1}$ Mpc$^{-1}$, $\Omega_{\Lambda 0}$ = 0.7, $\Omega_{m0}$ = 0.3. The redshift of NGC 6109 corresponds to a luminosity distance of 123 Mpc and a projected linear scale of 0.59 kpc arcsec$^{-1}$.

\section{VLA data calibration and analysis}

NGC 6109 was observed in S band (1.99 - 3.89 GHz) in August 2015 (project ID 15A-422) for 6000 seconds using the VLA in its A configuration and can be seen in Figure~\ref{Fig:6109}. For all observations, 3C 286 served as the primary flux density and polarisation position angle calibrator. The secondary calibrator 1635+3808 was used to monitor instrumental and atmospheric amplitude and phase fluctuations. The data were reduced using the Common Astronomy Software Applications package (CASA). The calibration closely followed the procedures set out in the CASA cookbook and reference manual. Data were `cleaned' \citep{hogbom} using multi-frequency synthesis to produce maps and all maps were self calibrated to maximise their dynamic range. Two Taylor terms were used to model the frequency dependence of the sky emission and spectral index maps were created from the ratio of the Taylor terms. Stokes IQUV were imaged so that the polarisation of the source could be investigated. We used the \textit{pyrmsynth} software to carry out rotation measure synthesis and create Faraday dispersion spectra within the components. This technique, introduced by \cite{burn} and extended by \cite{brentjens}, can overcome problems associated with the traditional linear $\phi / \lambda^2$ rotation measure relationship. These problems include the $n \pi$ ambiguity and rotation measure (RM) contributions from multiple components along the line of sight. \textit{pyrmsynth} incorporates the RMCLEAN method described by \cite{heald} and uses FFTs for Fourier inversion. The software takes cubes of Stokes Q and U images in frequency channels, and outputs cubes of Stokes Q and U in planes of constant RM. We averaged 50 channels (with width 2 MHz) per frame and `cleaned' the data to produce an image with 20 different frequency samples, ranging from 2.05 GHz to 3.95 GHz at 100 MHz intervals. RM synthesis was carried out with a weighting function of 1/$\sigma^2_{\textit{rms}}$ and 500 clean iterations. This gives a maximum resolution of 94.0 rad m$^{-2}$ and a maximum observable scale of 544 rad m$^{-2}$ in RM space.

\section{\chandra\ observations and reduction}
\label{sec:xobs}%2

We have used our earlier \chandra\ observations to measure the external pressures for comparison with those in the radio components. \source\ was positioned at the nominal aimpoint of the back-illuminated S3 chip of the \chandra\ Advanced CCD Imaging Spectrometer (ACIS) in VFAINT full-frame data mode on 2003 September 9th (OBSID 3985) and observed for 20 ks.  For the results here we reduced the data using {\sc ciao v4.9} with the {\sc caldb v4.7.3} calibration database, following the latest procedures described in the `threads' from the \chandra\ X-ray Center (CXC) \footnote{http://cxc.harvard.edu/ciao}.  These procedures apply the energy-dependent sub-pixel event repositioning algorithm. We masked unrelated sources using output from the {\sc ciao wavdetect} task.  

As originally discovered by \cite{feretti95} using \rosat\ data, the X-ray environment is complex, with \source\ embedded in a relatively poor cluster environment.  With the improved point spread function (PSF) of \chandra\ the emission from \source\ itself is separable into an AGN component and galaxy atmosphere (\citealp{evans2005}; \citealp{sun07}).  Because cluster emission covers the S3 chip we used the blank-sky fields to sample background for spatial analysis, filtering the source data for times of flaring using the same criteria as applied to the background fields, and re-normalizing to match rates at 9.5--12 keV where the particle component of the background dominates.  This cleaning shows the \chandra\ data for \source\ to have been subject to more background flaring than is usual, resulting in just 10.5 ks of data acceptable for use with the blank-sky data.  We are able to use local (cluster) background for fitting spectral models to the galaxy-scale gas, and here we have relaxed the flare screening criteria to 3$\sigma$ clipping, giving access to 19.0 ks of data.  All spectral fitting fixes absorption from gas along the line of sight in our Galaxy with $N_{\rm H} = 1.47 \times 10^{20}$ cm$^{-2}$. 

\section{Results}

\subsection{Large scale radio morphology}

\begin{figure}
  \includegraphics[width=8cm]{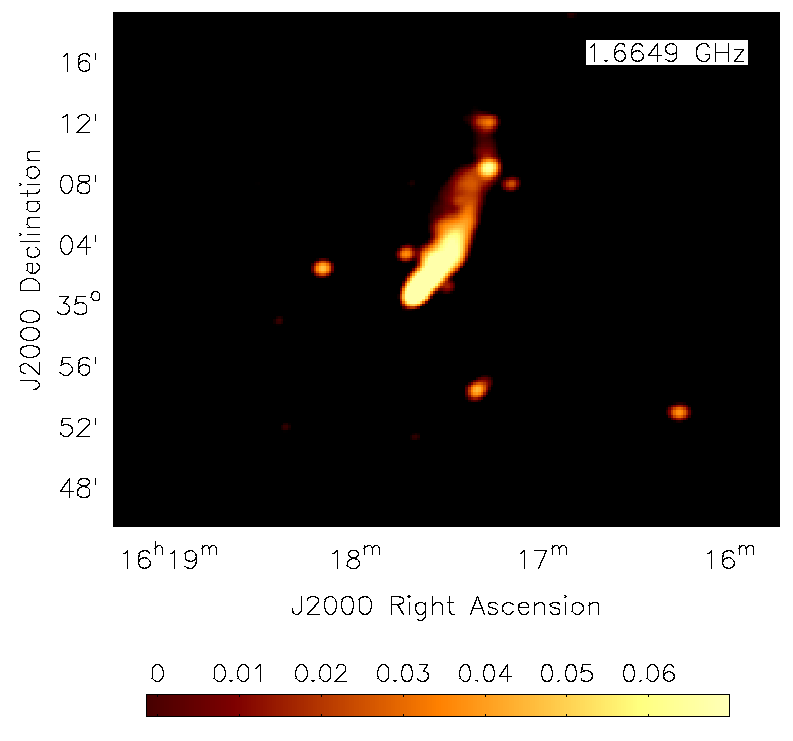}\\
  \caption{Radio image from the archival BT 7 project at 1.66 GHz and 44 arcsec resolution showing the extended radio structure of NGC 6109. The tail extends to the NW for a projected distance of 250 kpc.}
  \label{Fig:bt7}
\end{figure}

The large-scale radio emission associated with NGC 6109 is displayed in Figure~\ref{Fig:bt7} using archival VLA data at 1.67 GHz. It shows a tail extending to the NW for a projected distance of 250 kpc. On the large scale the emission looks similar to the head-tail source NGC 7385, except that there is no visible bifurcation of the tail \citep{rawes15}. A discrete background source at RA = 16$^h$ 17$^m$ 20$^s$, Dec = 35$^{\circ}$ 09$'$ is superimposed on the tail itself, and a weaker background source lies at the end of the tail. Emission along the tail remains bright for about 50 kpc, beyond which it rapidly drops off. The emission remains largely collimated to the N out to 100 kpc, and beyond this point the tail bends to the W and then to the N. The total flux density for this source is 1.87 Jy at 1.67 GHz. A radio source associated with NGC 6107 is visible SW of the core at RA = 16$^h$ 17$^m$ 20$^s$, Dec = 34$^{\circ}$ 54$'$. 

\subsection{Small Scale Structure}

\begin{figure*}
 \includegraphics[width=17cm]{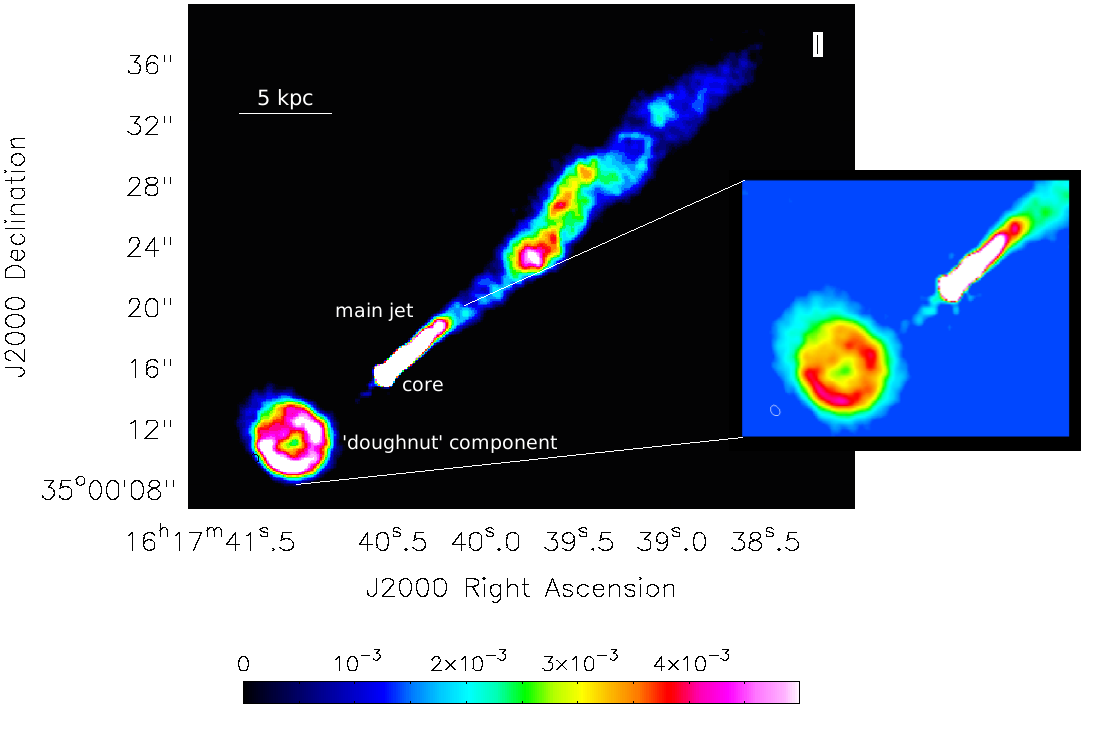}\\
  \caption{VLA S-band radio image of NGC 6109 at 0.8 arcsec resolution at 2.99 GHz. The main jet extends to the NW, and a bright knot is located approximately 12$''$ from the core. 4$''$ SE of the core, a tight swirl of radio emission is seen. This approximately circular component has a radius of 5$''$ and diffuse emission is observed to the NE. The colour bar shows the intensity scale in units of Jy beam$^{-1}$.}
  \label{Fig:6109}
\end{figure*}

The intensity distribution of NGC 6109 at 2.99 GHz is shown in Figure~\ref{Fig:6109} and morphological parameters determined from our observations are given in Table~\ref{Tab:newpara}. NW of the core the jet is collimated in the direction of the tail. The jet remains bright for 5$''$ before fading. The opening angle of the jet appears to increase within the fainter region. The jet brightens again approximately 12$''$ from the core, and remains bright for a further 12$''$. Within this high-emissivity region the emission bends to the N and then to the W and bright non-axisymmetric substructures are observed. 

A faint counter-jet is visible SE of the core. This extends 5$''$ before entering a tight swirl of bright radio emission. This swirl has a radius of 5$''$, and the emission at the centre of the component is noticeably fainter, giving it a doughnut-like appearance. The highest surface brightness is detected along the SE of the component, and there is broader low-surface brightness radio emission to the NE. The flux density enclosed within the swirl region is 21.0 mJy at 2.99 GHz. 

\begin{table}
\begin{centering}
\begin{tabular}{| c | c |} 
 \hline

Core flux density (mJy, 2.99 GHz) & 38.9 $\pm$ 0.04 \\
Core spectral index $\alpha$ & 0.05 $\pm$ 0.01 \\
Core rotation measure (rad m$^{-2}$) & 24 $\pm$ 1 \\
Core - SE component projected distance (arcsec) & 5.3 \\
SE component size (arcsec) & 5.3 $\times$ 5.6 \\
SE component flux density (mJy, 2.99 GHz) & 21.0 $\pm$ 0.5 \\
SE component spectral index $\alpha$ & 0.53 $\pm$ 0.02 \\
NW tail projected length (arcsec) & 43  \\
NW jet projected length (arcsec) & 5.9 \\
NW jet flux (mJy, 2.99 GHz) & 16.3 $\pm$ 0.4 \\
NW jet spectral index & 0.52 $\pm$ 0.02 \\
NW jet rotation measure (rad m$^{-2}$) & 32 $\pm$ 2 \\
Core - NW knot projected distance (arcsec) & 12 \\
NW knot projected length (arcsec) & 13 \\
NW knot flux (mJy, 2.99 GHz) & 24.7 $\pm$ 0.4 \\

\hline
\end{tabular}
\caption{Morphological parameters of NGC 6109 from the VLA observations. The spectral index $\alpha$ is defined as $S \propto \nu^{-\alpha}$.}
\label{Tab:newpara}
\end{centering}
\end{table}

%\begin{figure}
%\includegraphics[width=8cm]{6109cj.png}
%\caption{Small scale radio emission from NGC 6109 at 1.4 arcsec resolution at 1.51 GHz, highlighting the counter-jet.}
%\label{Fig:6109cj}
%\end{figure}

The core is found to have a spectral index, $\alpha$, defined in the sense $S_{\nu} \propto \nu^{-\alpha}$, of $\alpha$ = 0.05 $\pm$ 0.01 and the spectral index of the main jet is $\alpha$ = 0.52 $\pm$ 0.02 along the first 6$''$. Further from the core the emission is too faint to give good results. In the swirl structure the spectral index is 0.52 $\pm$ 0.02 in agreement with the main jet and there is little structure in the spectral index across the component. These values agree with the results of \cite{ekers78}, who found the inner jet to have $\alpha$ = 0.5. Ekers also reported $\alpha$ = 1 along the tail out to 50 kpc and more pronounced steepening of the spectral index at higher frequencies, suggesting the existence of a break in the radio spectrum. 

\subsection{Polarisation and Rotation Measure Structure}

The distribution of polarised radio emission in NGC 6109 is reported by \cite{fanti81} out to 100 kpc. At 1.4 GHz, they find that the polarisation increases along the jet, from 6\% in the first few bright regions, up to 20\% in the extended tail. They also report significant depolarisation between 5 GHz and 1.4 GHz in the `blob' regions, approximately 20, 50, 60 and 100 kpc away from the core. 

Our VLA data reveal the polarisation structure of the inner $\sim$ 20 kpc at high resolution and probe the magnetic field configuration. We used Stokes I, Q and U images at 0.8 arcsec resolution to derive the distributions of degree of polarisation and polarisation position angle at 2.99 GHz. The degree of polarisation was corrected to first order for Ricean bias \citep{wardle74}.

\begin{figure}
\begin{centering}
 \subfigure{\includegraphics[width=9cm]{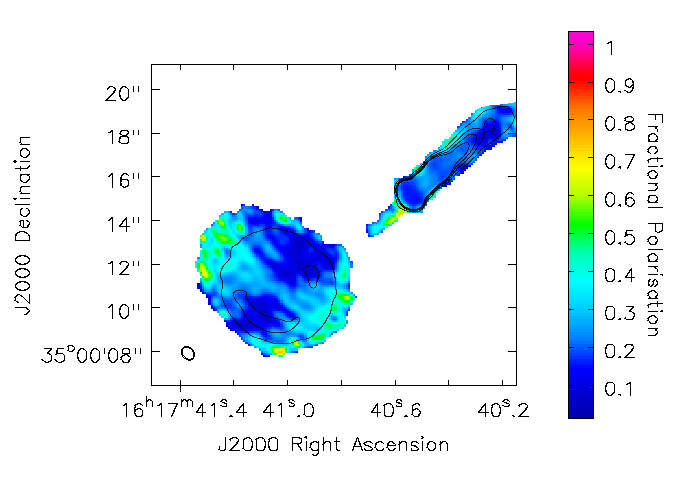}}
  \hfill
  \subfigure{\includegraphics[width=9cm]{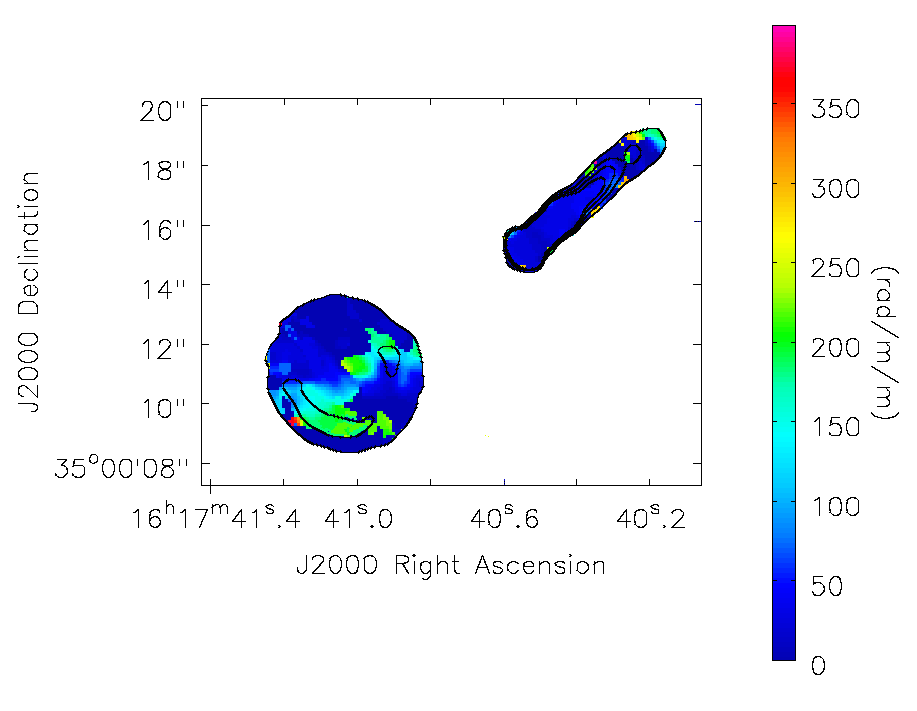}}
  \caption{\textit{Top:} The fractional polarisation (beam in bottom left) and \textit{bottom:} the RM structure in the doughnut component and along the main jet. Both images are shown with VLA total intensity contours at 0.16, 0.32, 0.48 and 0.64 mJy beam$^{-1}$.}
  \label{Fig:fpol1}
\end{centering}
\end{figure}

A fractional polarisation map is given in Figure~\ref{Fig:fpol1}. To minimise errors in measurements of $P$ caused by residual sidelobes in the total intensity data, all pixels with errors $>$ 0.01 in fractional polarisation are blanked. The peak fractional polarisation in the image is 51\%, similar to the peak values found in other radio sources \citep{bridle84} and comparable to the theoretical maximum for a source with a uniform magnetic field \citep{pacholczyk70}. In the core, the polarisation is $P$ = 0.11 $\pm$ 0.02. Along the bright part of the main jet $P$ = 0.15 $\pm$ 0.02. In the doughnut component, the SW and NE sides have fractional polarisation $P$ = 0.31 $\pm$ 0.03, whereas the SE and NW sides display lower polarisation $P$ = 0.09 $\pm$ 0.02. The doughnut shows an increased fractional polarisation around its edge, as found in radio source lobes (e.g. \citealp{laing96}). 

%\begin{figure}
%\includegraphics[width=8cm]{fpol1}
%\caption{Fractional polarisation structure of NGC 6109 at 1.51 GHz. Radio contours are shown to highlight the edges of the swirl region at 0.16mJy, 0.32mJy, 0.48mJy and 0.64mJy}
%\label{Fig:6109fracb}
%\end{figure}

%\begin{figure}
%\includegraphics[width=8cm]{phi}
%\caption{Rotation Measure structure in NGC 6109 at 1.51 GHz produced via rotation measure synthesis. Radio contours are shown to highlight the edges of the swirl region at 0.16mJy, 0.32mJy, 0.48mJy and 0.64mJy}
%\label{Fig:6109rm}
%\end{figure}

\begin{figure}
\begin{centering}
 \subfigure{\includegraphics[width=8cm]{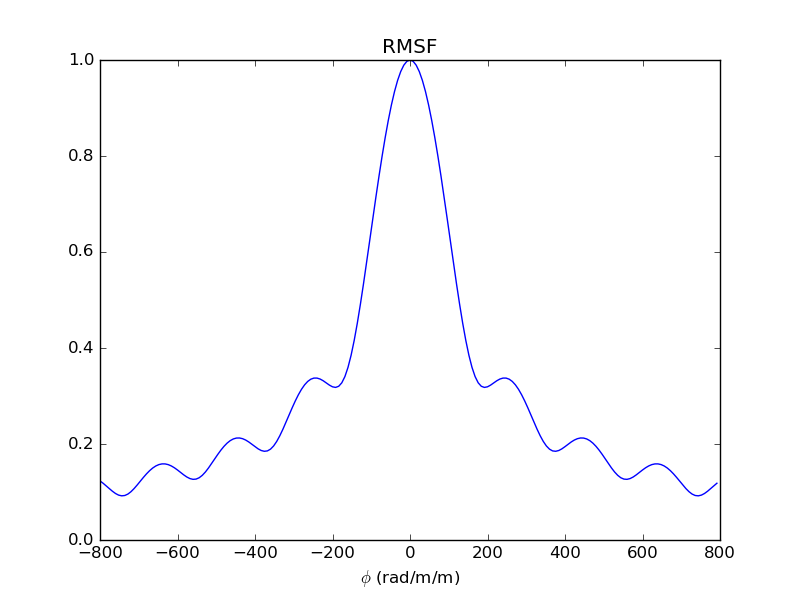}}
  \hfill
 \subfigure{\includegraphics[width=8cm]{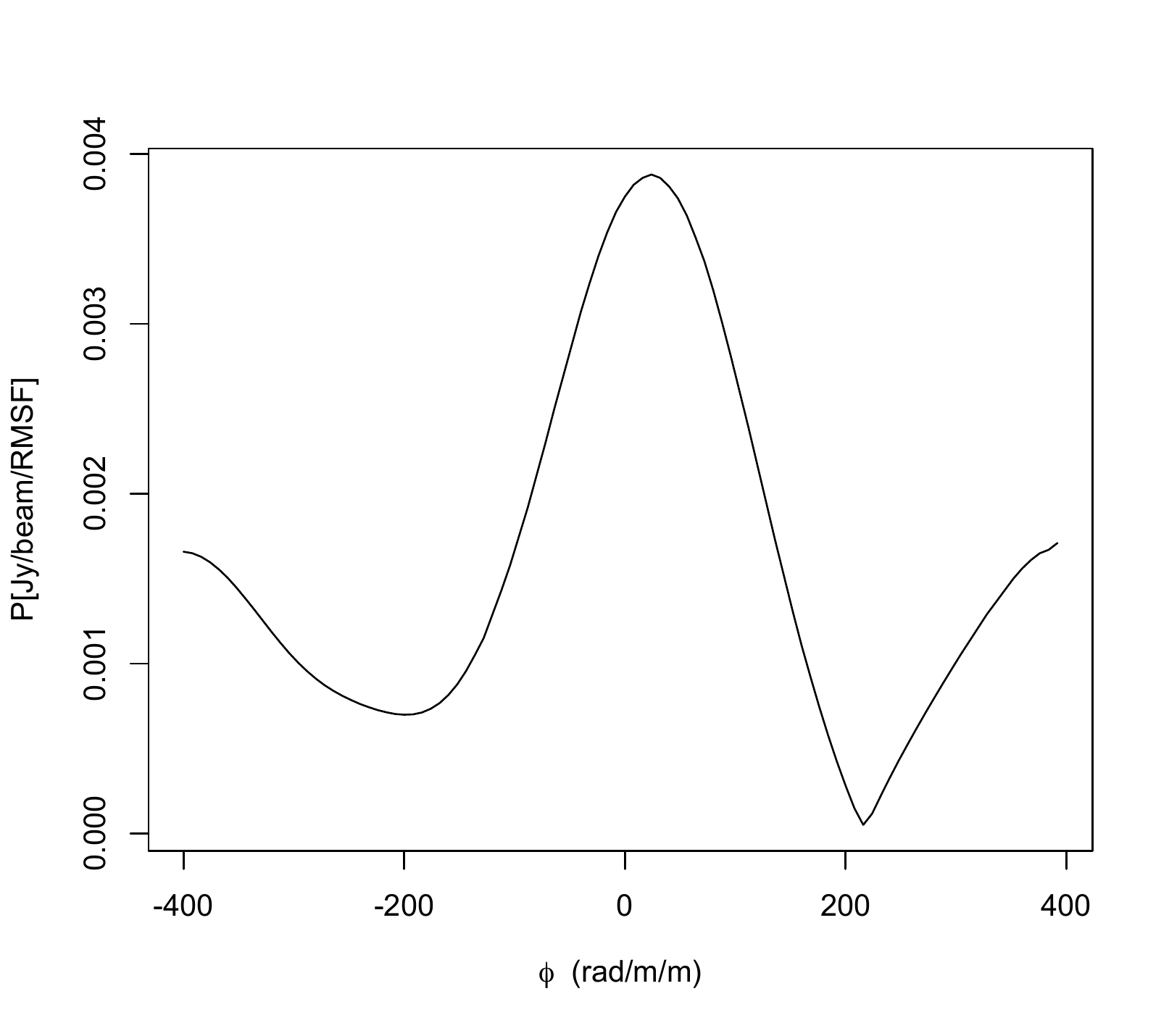}}
  \hfill
  \subfigure{\includegraphics[width=8cm]{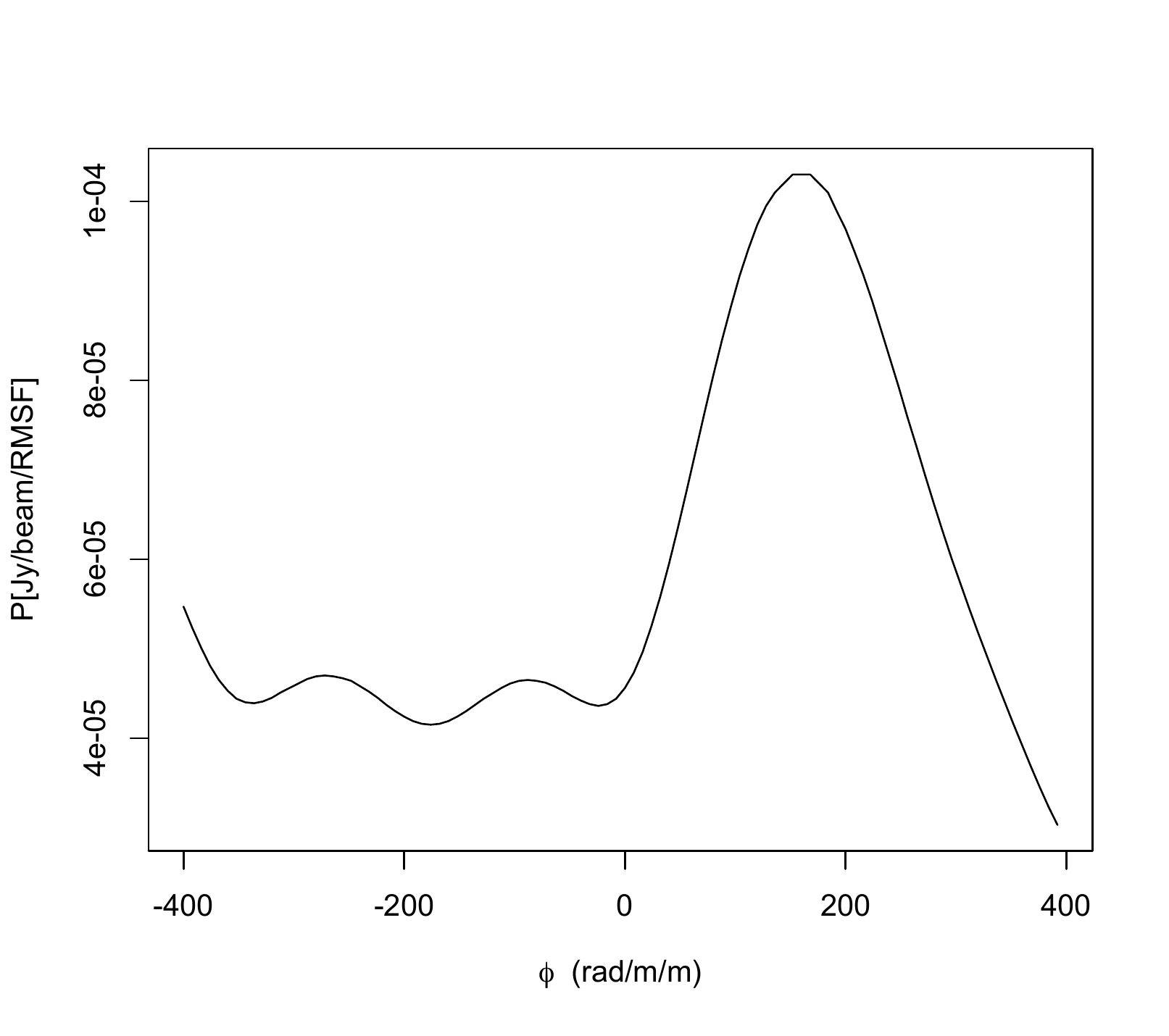}}
  \caption{\textit{Top:} The RMSF for the RM synthesis code as applied to NGC 6109 for frequency range 2.05 GHz to 3.95 GHz with 100 MHz intervals. The Faraday dispersion distribution in the core and southern region of the doughnut component of NGC 6109 are given in the middle and lower figures, respectively.}
  \label{Fig:core1}
\end{centering}
\end{figure}

Rotation measure synthesis was used to investigate the doughnut component and the main jet.  The rotation measure spread function (RMSF) is given in Figure~\ref{Fig:core1}. The fitted FWHM of the clean beam is 94 rad m$^{-2}$. In the core, the RM is found to be 24 $\pm$ 1 rad m$^{-2}$ in agreement with \cite{taylor09}. Along the main jet, the rotation measure is constant, at 32 $\pm$ 2 rad m$^{-2}$ for 5$''$. Beyond this region the RM is unavailable due to low signal to noise, until the bright knot at 12$''$ from the core. Within the knot, the rotation measure is 40 $\pm$ 8 rad m$^{-2}$.

Figure~\ref{Fig:fpol1} shows the RM synthesis peak values in the doughnut component with 2.99 GHz VLA total intensity contours. The regions of high RM in the doughnut appear to correspond to regions of high radio flux density and we find no significant polarised structure below 0 rad m$^{-2}$. Faraday dispersion spectra in the core and in the highest flux region of the doughnut are given in Figure~\ref{Fig:core1}. The sidelobes on these spectra are low, and the observed peaks are in good agreement with values obtained by traditional $\phi / \lambda^2$ fitting.

Regions of high RM in the doughnut are correlated with regions of low fractional polarisation. It is probable that the emission has been depolarised by fine structure in the Faraday screen responsible for the RM, or in the doughnut itself. These mechanisms could be distinguished by future investigations of the RM and emission structure on sub-beam scales, with lower frequency LOFAR observations or higher-frequency VLA mapping. Higher fractional polarisation is detected in the NE-SW direction where the RM is similar to the main jet, at 32 rad m$^{-2}$. The RM along the main jet has a uniform structure, suggesting that this RM is associated with a foreground screen outside the galaxy atmosphere. The RM in the doughnut shows as a stripe (along an extension of the counter-jet) and a `cap' at the SE edge of the doughnut, suggesting the foreground screen on this side is created by the jet (and counter-jet) itself. 

\begin{figure*}
\begin{centering}
 \subfigure{\includegraphics[width=8cm]{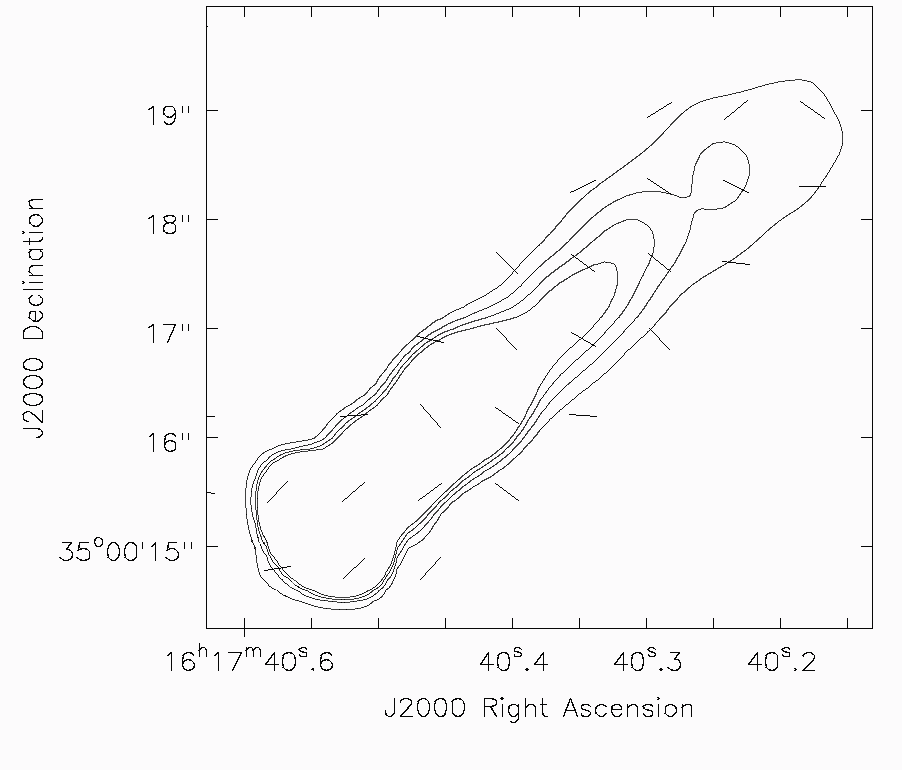}}
  \hfill
  \subfigure{\includegraphics[width=8cm]{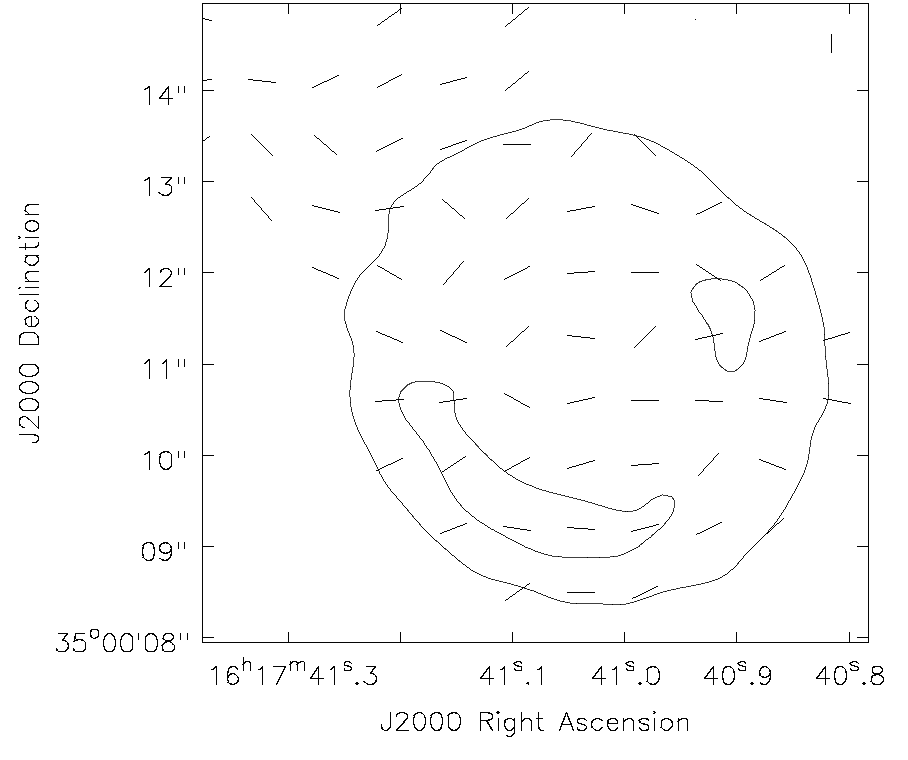}}
  \caption{The apparent magnetic field direction in the (\textit{left}) core and along the first 5 kpc of the main jet and (\textit{right:}) in the doughnut component. Both images are shown with VLA total intensity contours at 0.16, 0.32, 0.48 and 0.64 mJy beam$^{-1}$.}
  \label{Fig:vectors1}
\end{centering}
\end{figure*}

Using the calculated RM values, we produced maps of the apparent intrinsic magnetic field direction. These maps are given in Figure~\ref{Fig:vectors1} for the brightest part of the main jet and the doughnut component, with 2.99 GHz VLA total intensity contours. In the core and main jet, the magnetic field is as expected for a low power FRI type jet; the vectors rotate from parallel to the jet direction to perpendicular further out. In the doughnut component however, the magnetic field shows a combination of circumferential and transverse components. In the bright southern region, the field is circumferential, as could be produced by shearing of the field into a direction tangential to the source boundary. Such structure is commonly observed in other sources where jet bending is evident as well as in radio lobes. At the centre of the component the field appears transverse to the direction of the main jet. Magnetic field to the NE, where diffuse emission is thought to arise from the jet expanding into the environment is mostly transverse, although longitudinal emission is seen on the E side, possibly at the boundary between the radio plasma and the external medium.

\section{X-ray results}
\label{sec:xresults}%

\begin{figure}
\centering
\includegraphics[width=0.85\columnwidth]{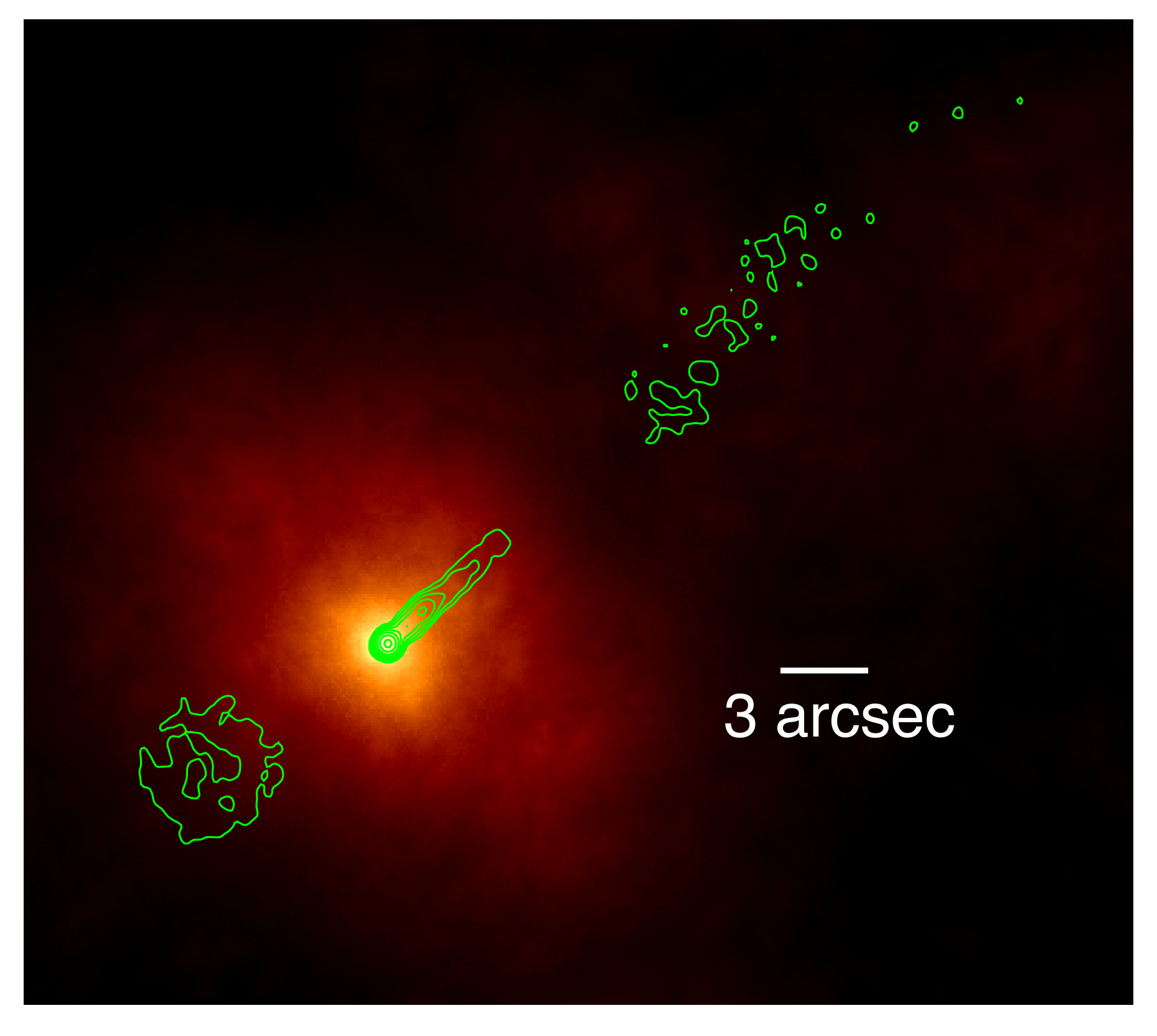}
\caption{Exposure-corrected 0.4-5 keV {\it Chandra\/} image in 0.0984-arcsec pixels adaptively smoothed with a circular top hat filter of 20 counts. The radio contours are of a combined A- and B-array 4.86-GHz VLA map from programme AH766. The beam size is 0.46 x 0.43 arcsec and the contours are in increasing factors of two from 0.15 mJy/beam,}
\label{fig:xrayzoom}
\end{figure}

From \rosat\ it was known that cluster X-ray gas encompasses the two brightest cluster galaxies \source\ and NGC\,6107, with an envelope of major axis in a roughly NE-SW direction \citep{feretti95}. The cluster emission sampled with {\it ROSAT\/} spatially fills the S3 chip of {\it Chandra\/}. It sits on top of a background, which is measured for radial-profile analysis as described in Section 3.  The combined cluster and background emission shows as sparse counts in the overall image and so cannot meaningfully be subtracted for image display. On the smaller scale an image showing the NGC 6109 galaxy emission (with cluster and background emission included) is given in Figure~\ref{fig:xrayzoom}. The galaxy atmosphere of \source\ has a similar orientation to the cluster gas, with its long axis orthogonal to the general lie of the radio features, as found in a significant fraction of nearby radio galaxies (Duffy et al. 2018, in preparation). While there may be a few excess counts associated with synchrotron radiation from the NW radio jet within about 3 arcsec of the core, it is clear from Figure~\ref{fig:xrayzoom} that no obvious X-ray features are associated with the counter-jet. In this work we use a spherical approximation to describe the gas distribution as this allows straightforward estimates of pressure, and because the extent of the emission along the line of sight is uncertain.

\begin{figure}
\centering
\includegraphics[width=0.85\columnwidth]{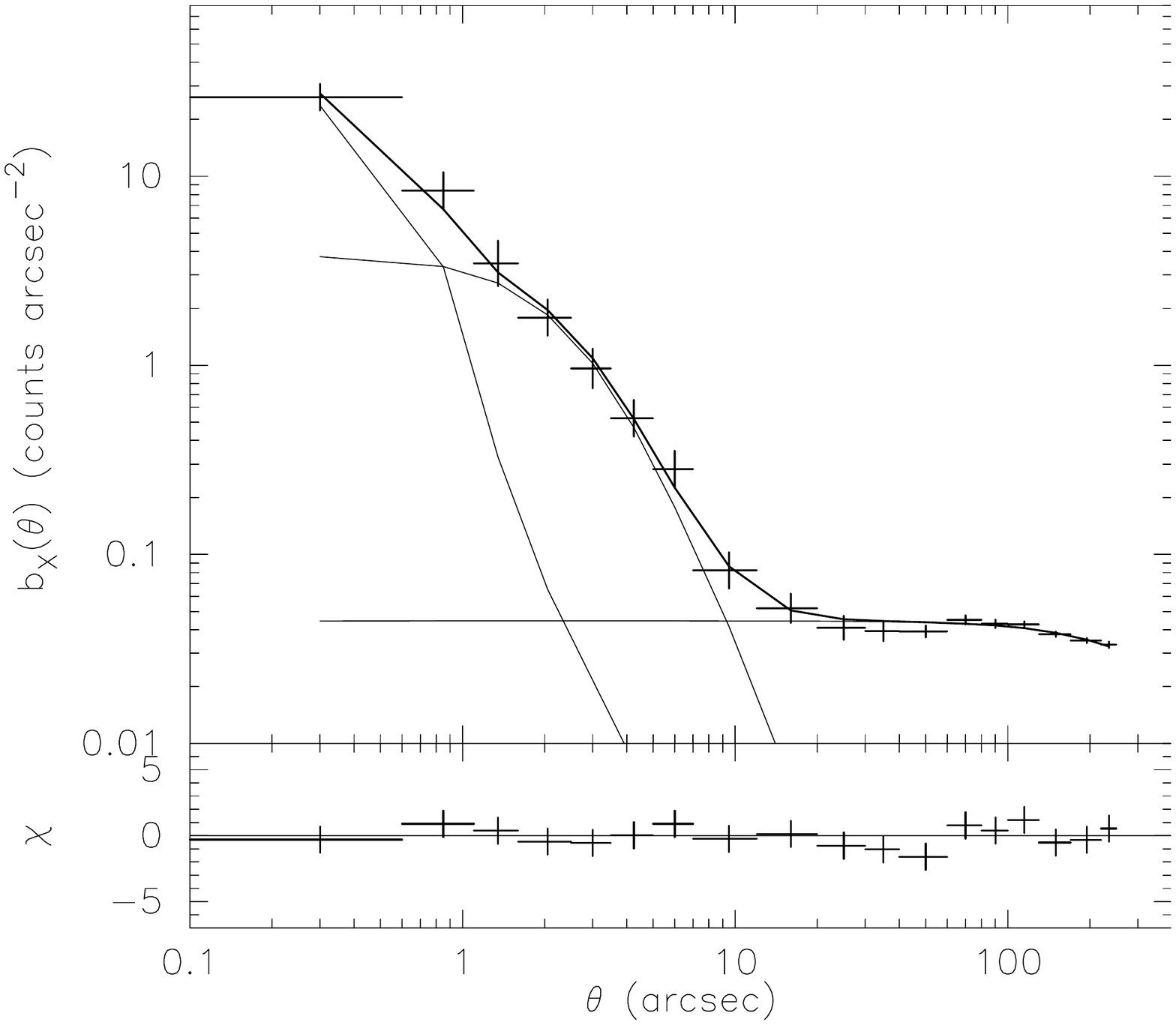}
\caption{ Background-subtracted exposure-corrected 0.4-5 keV radial profile. The broadest model component describes the cluster gas.  The narrowest component is the PSF, representing the AGN core, and the final component is a $\beta$-model (density $\propto (1 + (\theta^2/\theta_{\rm c}^2))^{-3\beta/2}$) convolved with the PSF, representing the galaxy-scale gas.  The galaxy component is best fitted with parameters $\beta=0.8$, $\theta_{\rm c} = 2.8$ arcsec ($\chi^2/$dof = 9.4/15).}
\label{fig:gasradial}
\end{figure}

Figure~\ref{fig:gasradial} shows the radial profile. The number of counts in the resolved galaxy emission within 10 arcsec of the galaxy nucleus is roughly twice that in the AGN.  Guided by the radial profile, we extracted counts from a core-centred annulus of radii 1.3 and 10 arcsec to sample the galaxy spectrum, with background from a core-centred annulus of radii 20 and 70 arcsec, and we fitted the result to a thermal (APEC) model.  Similar results for the temperature were found by fitting all the emission in a circle of radius 10 arcsec to the combination of a power law and thermal emission.  Abundances were poorly constrained but showed a preference for low values and were subsequently fixed at 20 per cent of Solar. Our fit gives $kT = 0.76^{+0.10}_{-0.12}$ keV (90\% uncertainties). This is consistent within uncertainties with the temperature found by \cite{evans2005} who made the first attempt to separate AGN and gas emission using the \chandra\ data. The gas temperature associated with the galaxy is significantly lower than the temperature deduced by \cite{feretti95} who fitted an optically thin plasma model to the cluster gas detected by the ROSAT satellite and found $kT = 2.4\pm$1.2  keV. This is likely due to a combination of spectral considerations, the choice of regions adopted to measure the gas properties and a larger uncertainty for $kT$ as measured by ROSAT.

We measure parameters of the gas including density and pressure as a function of radius using the spatial and spectral results and methodology presented in \cite{birkinshaw93} and \cite{worrall2006}. We find a total bolometric luminosity for the galaxy component of $(1.4^{+0.3}_{-0.2}) \times 10^{41} $ ergs s$^{-1}$, in excellent agreement with \cite{sun07}. 

\section{Discussion}

NGC 6109 was originally classified as a `head-tail' type galaxy by \cite{colla75}, based on radio observations from WSRT at 50 cm. \cite{odea85} re-classified it as a narrow angle tailed (NAT) source. NAT morphology is attributed predominantly to the interaction of radio emission ejecta by a moving galaxy with the intracluster medium (e.g. \citealp{jones79}; \citealp{christiansen81}). The bending of the radio jets in NAT's provides important constraints on the physical conditions within the jets and the immediate environment in the galaxy and intracluster medium. 

A faint radio counter-jet is visible in Figure~\ref{Fig:6109}, SE of the core. The expression

\begin{equation}
R = \left(\frac{1+ \beta \cos\theta}{1 - \beta \cos\theta}\right)^{2+\alpha}
\end{equation}

\noindent gives $R$, the ratio of the brightness between the jet and the counter-jet where $\alpha$ is the spectral index, $\beta$ is the bulk speed relative to the speed of light and $\theta$ is the angle between the line of sight and the jet for relativistic outflows. This expression follows the assumptions described in detail by \cite{laing13} for modelling intrinsically symmetrical, axisymmetric, decelerating relativistic outflows. At 1.8$''$ from the core, we find $\alpha$ = 0.52 $\pm$ 0.02 and $R$ = 17.0 $\pm$ 0.3, constraining $\beta \cos \theta$ to 0.51 $\pm$ 0.02. This implies $\cos \theta >$ 0.5, thus $\theta <$ 60$^{\circ}$. The expression $D$ =  (1 + $\beta \cos \theta$) / (1 - $\beta \cos \theta$) gives the corresponding Doppler distance ratio, $D$ and we find $D$ = 3.1 $\pm$ 0.2. The SE component at a projected distance of 4$''$ from the core, if moving with the same bulk speed as the jet, would be associated with a two-sided ejection event that would produce a matching feature 12.4$''$ along the main jet. Looking at Figure~\ref{Fig:6109}, the bright knot along the main jet lies 12$''$ from the core, suggesting that the flaring region could be an outflow feature emitted at the same time as the SE component, in the opposite direction. 

We calculated the equipartition energy density using the synchrotron code of \cite{hardcastle98}. We assumed equipartition between the electrons and the magnetic field and negligible relativistic beaming. An electron spectrum with $\gamma_{\text{min}}$ = 10 was used, where $\gamma$ is the Lorentz factor of the electrons ($\gamma$ = $E / m_e c^2$). For the doughnut region with radius 5$''$ and an estimated length along the line of sight of 3$''$, the equipartition energy density is U$_{\rm eq}$ = 9.1 $\times$ 10$^{-12}$ J m$^{-3}$. The NW knot on the main jet side was modelled with a length 20$''$ and radius 1$''$, to include the full extent of the flaring region as indicated by Figure~\ref{Fig:6109}. This gives U$_{\rm eq}$ = 9.5 $\times$ 10$^{-12}$ J m$^{-3}$. The similarity of these energy densities supports the idea that the knot is the counterpart of the doughnut, ejected in the opposite direction. Relativistic effects on the energy density are considered to be minimal for $\beta$ = 0.5. The magnetic field strength in the SW component is calculated as $B_{\text{min}}$ = 3.2 nT and the minimum pressure $P_{\text{min}}$ = 3.0 $\times$ 10$^{-12}$ J m$^{-3}$. The external pressure in X-ray emitting gas measured from the galaxy component of Figure~\ref{fig:gasradial} at 10$''$ from the core is found to be $P = 1.4 ^{+0.5}_{-0.5} \times 10^{-12}$ J m$^{-3}$ (90\% uncertainty). The galaxy gas density is calculated as $n = 5.1 ^{+1.0}_{-1.6} \times 10^{-3}$ cm$^{-3}$. 10$''$ is the central projected angle for the location of the doughnut. For $\theta <$ 60$^{\circ}$ the doughnut will lie further out in the atmosphere where the external pressure will be lower.  

While we estimate that the inner radio jet and doughnut may be a factor $\geq$ 3 over-pressured, pressure equilibrium between ambient gas and the radio emission in the main jet overall is reported by \cite{feretti95}. However they find that further along the tail, about 100 kpc from the core, there is a larger apparent imbalance towards the external pressure dominating. This is consistent with the findings of \cite{killeen88} and \cite{feretti92}. They argue that it is likely that assumptions used for the calculation of equipartition parameters cause the appearance of imbalance and that the entire tail is confined by the intergalactic medium. They suggest that the energy ratio between relativistic protons and electrons may be between 20-60, rather than unity, or that there may be a population of electrons that would radiate below the low frequency limit of the observable spectrum. If this is true in the doughnut and inner jet then these radio components will be highly over-pressured. 

The unique shape and extreme distortion of the radio emission require an interpretation that can account for the morphology, polarisation structure and rotation measure presented in the previous sections. Helical motion within jets on a parsec and kpc scale has been modelled in both magnetohydrodynamic and force-free jet models (\citealp{devilliers05}; \citealp{mckinney06}; \citealp{mckinney09}) and these models have sufficient resolution to compare structure with features observed in AGN jets (e.g. \citealp{hardee01}, \citealp{hughes02}). \cite{hardee03} note that the polarisation in helically modelled jets is typically well below the maximum value of 70\% for synchrotron radiation. They attribute this to the magnetic field in jets beyond the acceleration region not being well organised. In NGC 6109 we also see polarisation of $\sim$ 20\% in the jets, which follows this pattern. 

\subsection{Signatures of helical jets}

Observational signatures of helical jets have been found in X-ray binaries (e.g. \citealp{hjellming95}), planetary nebulae (e.g. \citealp{lopez93}) and the jets of AGN. The term `helical jet' can be used to describe at least three different morphological jet structures and these are detailed by \cite{steffen97}. Firstly, ballistic helical jets describe jets in which the individual jet fluid elements flow along straight lines but the overall structure is helical due to the periodic change in the ejection direction of the elements. These jets are described as precessing. Secondly, helically bent jets have fluid elements that flow along a common twisted path, delineated by the curved jet axis. Jets can be bent by several mechanisms, such as transverse winds, collisions with gas clouds, density gradients or a dense ambient medium (e.g. \citealp{ludke94}. Finally, jets with an internal helical structure are straight as a whole, but have fluid flowing along helical trajectories within the jet. Helical trajectories can also be caused by Kelvin-Helmholtz instabilities (e.g. \citealp{hardee87}; \citealp{hardee92}; \citealp{conway93} and  hydrodynamical (e.g. \citealp{zhao92}; \citealp{hardee94}) simulations have been carried out to investigate how this can disrupt the path of a jet. \citet{koide96} and \citet{nishikawa99} pioneered magnetohydrodynamic studies of how a strong, ordered magnetic field can affect jet propagation. \citet{koide96} found that the bending scale depended on both jet velocity and jet magnetic field angle. Slower jets with an angle between the jet and magnetic field of 45$^{\circ}$ were bent the most. 

An important property of helical jet structures is that the observed properties are not always symmetric with respect to the axis of the helix. The proper motion and magnetic field vectors on one side of the helix will point in a direction closer to the line of sight than those on the opposite side. This means that quantities such as the flux, optical depth, polarisation and rotation measure can be asymmetric, especially if the speed of the jet is relativistic.

There is no evidence in the local environment of NGC 6109 for collisions with gas clouds or for significant density enhancements in the galaxy or cluster X-ray emitting atmospheres that relate to structures in the radio source. The existence of a high RM along the southern edge of the doughnut component, however, suggests the presence of magnetised plasma interacting with or wrapped around the radio object. This magnetised plasma could be the remnants of a radio trail from a galaxy that has passed close to NGC 6109. The brighter radio surface brightness in this region could suggest shock compression, where the radio emission has encountered magnetised gas and been compressed, resulting in jet deflection. Although a viable possibility, this hypothesis doesn't explain the size of the jet bend, $>$ 180$^{\circ}$. In the following sections we discuss both precessing and helically bent jet models.

\subsection{Ballistic model}

A ballistic model of a helical jet, based on jet precession, can explain the doughnut-like feature if the angle to the line of sight is small. The component can then be interpreted as a helical outflow, viewed almost end on so that the emission appears approximately circular. This scenario however, offers no clear explanation of the high regions of rotation measure detected in the south of the component.

\begin{figure}
\begin{centering}
\includegraphics[width=8cm]{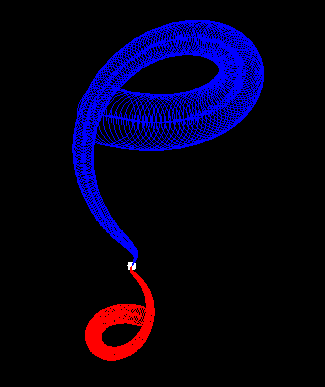}
\caption{A ballistic simulation shows the shape of precessing jets for $\beta$ = 0.5, $\theta$ = 20$^{\circ}$ and other parameters described in the text. The intensity contrast is not modelled, and the blue trajectory is directed towards the observer and the red trajectory is directed away. In this representation, NS orientation is arbitrary.}
\label{Fig:ballistic}
\end{centering}
\end{figure}

\cite{steenbrugge08} use a ballistic model to interpret the radio emission in Cygnus A. They find that the radio knots that delineate the jet and deviate from a straight line can be satisfactorily fitted with the precession model of \cite{hjellming95}. In this model, symmetric jets are launched along an axis which traces a cone throughout a precession period $P$, of opening angle $\phi$ and inclined to our line of sight at angle $\theta$. We used a similar ballistic model to see whether we could replicate the shape of the observed structure of NGC 6109. We adopted a jet speed of $\beta$ = 0.5, based on the jet/counter-jet ratio calculations with $\theta$ taken to be small, to reproduce the doughnut feature. Figure~\ref{Fig:ballistic} gives a simulation which is able to produce a doughnut-like component, however a looped jet is also ejected on the opposite side of the core. Here $\phi$ = 10$^{\circ}$ and $\theta$ = 20$^{\circ}$. The jet half opening angle is 2$^{\circ}$. The model has a jet precession period of $T_j$ = 1.8 $\times$ 10$^4$ years and reproduces the observed structure reasonably well except for a looping structure in the approaching jet (blue) that is not seen in the data. This blue loop must be disrupted in some way to become non-ballistic at a certain distance from the core, whilst the red loop is unaffected. This could be due to the transition from the galaxy to cluster external atmosphere at $\sim$ 10$''$, however this would require an angle to the line of sight $\theta >$ 35$^{\circ}$ in order to keep the red loop within the region where the galaxy atmosphere dominates. 

This ballistic model doesn't include effects such as deceleration of the jet as it propagates beyond a few kpc, and should therefore only be used as a guide to whether such loops can be produced ballistically. Precession models by \cite{gower82} highlight how the beam velocity can cause significant geometry changes in twin relativistic jets. In addition, \cite{gower82} show that at lower values of $\beta$ the swirls can be tighter but the two-sided structures are more symmetric, in contradiction to what is observed for NGC 6109. 

\subsection{Possible causes for precession}

Is it not clear why the jet launch direction would be precessing. A few mechanisms that have been suggested are described below.

Evidence for central engines of AGN formed by massive binary systems includes double nuclei (NGC 4486B, \citealp{lauer96}), wiggly jets (e.g. \citealp{kaastra92}) and periodic optical light curves (OJ 287, \citealp{villata98}). Radio structure associated with the AGN, built up over many precession periods, will be aligned with the orbital angular momentum. On the sub-kpc scales observed by the VLA however, precession related curvature could be discernable. In this mechanism, the spin axes will undergo geodetic precession about the total angular momentum. For the more massive black hole the precession period P$_{\text{prec}}$ in years is given by \cite{begelman80} as

\begin{equation}
P_{\text{prec}} = 600 r_{16}^{5/2} \left( \frac{M_8}{m_8} \right) M_8
\end{equation}

\noindent where $r_{16}$ is the distance in units of 10$^{16}$ cm and $M_8$ and $m_8$ are the masses of the larger and smaller black holes in units of 10$^8$ $M_{\odot}$.

Using the jet precession period of 1.8 $\times$ 10$^4$ years, as adopted for Figure~\ref{Fig:ballistic}, a black hole mass $M_8$ = 1 and binary mass ratio of 10 gives an orbital radius of order 5 pc. This corresponds to a wide binary system, where the precession time scale should be comparable with the inferred lifetime of extended radio components. The orbital time period for the binary system is given by

\begin{equation}
P_{\text{orb}} = 1.6 r_{16}^{3/2} M_8^{-1/2} \rm{yr}
\end{equation}

\noindent assuming Keplerian motion. For $r$ = 5 pc this gives $P_{\text{orb}}$ = 600 yrs. Evidence for a binary system could be provided from the detection of double or displaced emission lines or by looking for proper motion of the AGN ($\sim$ 1 $\mu$as yr$^{-1}$) with multi-epoch VLBI.

The Lense-Thirring effect can also cause jet precession. This effect is the frame dragging produced by a rotating compact body and causes precession of a particle's motion if its orbital plane is inclined to the equatorial plane of the rotating object \citep{lense18}. The precession angular velocity $\Omega_{\text{LT}}$ is given by (e.g. \citealp{wilkins72})

\begin{equation}
\Omega_{LT}(R) = \frac{2 G}{c^2} \frac{J}{R^3}
\end{equation}

\noindent where $J = aGM^2/c$ is the black hole angular momentum and $a$ is the black hole spin parameter. For a precession period T$_{\text{prec}}$ = 1.8 $\times$ 10$^4$ years, a black hole mass in the range 1 $\leq$ M$_8$ $\leq$ 100 and a spin parameter $a$ = 0.1 - 0.9, the orbital radius $R$ is constrained to 1 $\leq R \leq$ 40 parsec.

The combined action of the Lense-Thirring effect and the internal viscosity of the accretion disk forces alignment between the angular momenta of the black hole and the accretion disk \citep{bardeen75}. This effect only concerns the innermost part of the disk due to the short range of the Lense-Thirring effect, while the outer part of the disk remains in its original configuration. \cite{scheuer96} showed that the disk alignment and precession time scales are identical. This mechanism has been invoked in cases of NGC 1097 \citep{caproni04b}, NGC 1068 \citep{caproni06b} and NGC 4258 \citep{caproni07}. The Bardeen Petterson transition radius between the spin-aligned and outer parts of a thin disk, based on the hydrodynamic simulations of \cite{nelson00}, can be expressed as 

\begin{equation}
R_{BP} = A a^{2/3} R_{g}
\end{equation}

\noindent where the scaling parameter A is a function of the viscosity of the radial accretion flow \citep{shakura73} and the aspect ratio of the disk, with 10 $\leq$ A $\leq$ 300. This suggests R$_{BP} >$ 1 pc for the NGC 6109 system. 

Based on the \cite{sarazin80} model of disk-driven precession for SS 433, it was suggested by \cite{lu90} that a tilted accretion disk can also produce jet precession. \cite{petterson77} showed that if an accretion disk is optically thick, radiation pressure can produce nonaxisymmetric torques that will change the initial configuration of the disk. The influence of magnetic fields on accretion disks has also been investigated (e.g. \citealp{lipunov80}; \citealp{terquem00}; \citealp{lai03}; \citealp{pfeiffer04}).

These binary and single AGN precession models however are unable to account for the observed magnetic field and rotation measure structure in NGC 6109. In the precessing system SS 433, the magnetic field is found to track the helical trajectory of the jet on a parsec scale (\citealp{hjellming81}, \citealp{stirling04}). If this magnetic field structure remains stable out to kpc scales during jet precession, then we would expect to see a circumferential field within the doughnut component. Although such a field is observed to the south of the component, the majority of the field structure in the component is radial. SS 433 also shows evidence for high fractional polarisation at the leading edges of the helices \citep{roberts08} produced as the jet material flows into an ambient medium containing a tangled magnetic field. The resulting compression orders the field in two dimensions \citep{bridle84}. NGC 6109 is not significantly polarised at the edges of the doughnut, highlighting the lack of large scale ordering of the magnetic field in this region. The presence of a non-uniform rotation measure across the component supports the hypothesis that some external magnetic material may be contributing to, or causing the large jet bending. The orientation of the band of high RM (i.e. anti-parallel to the main jet) however, suggests that it is produced by the counter-jet. Since the doughnut is the more distant structure, the counter-jet is superimposed on top of it. 

\subsection{Jet deflection models}

A significant number of radio galaxies have been reported to display jet bending on kpc scales. Parallels between NGC 6109 and NGC 7385 can be drawn, as both have been categorised as low redshift `head-tail' galaxies by the 3CRR survey \citep{laing83} and exhibit $\geq$180$^{\circ}$ jet deflections. A swirl-like feature has also been observed for the FRI-type radio galaxy NGC 7016 (\citealp{cameron88}; \citealp{worrall14}), though this lies substantially further from the host galaxy. In the case of NGC 7385 our Hubble Space Telescope (HST) data revealed a large optical cloud in the path of the counter-jet, and the interaction between the jet and the cloud is believed to cause the strong jet bending (Rawes et al, in prep). Unfortunately NGC 6109 has not been observed with HST, however infrared and optical observations from \textit{Spitzer}, CFHT and KPNO show no similar deflector in the vicinity of the counter-jet component. Therefore although both objects exhibit significant jet bending, different interaction types are required to explain the deflections. 
Less extreme jet bending has been observed in other FR I-type galaxies (e.g. 3C 66B, \citealp{hardcastle96}; 3C 321, \citealp{evans08b}), and it is possible that the mechanisms which produce these deflections could also apply to NGC 6109. Various proposals for bending jets include buoyancy or ram pressure in a dense ICM \citep{miley72} and orbital motion (\citealp{blandford78b}; \citealp{roos1993}).

\cite{eilek84}, studying the wide-angle tail source 3C 465, modelled the C-symmetric source in several bending scenarios. Since the jets within the host galaxy are straight, all models ascribe the effect to the ICM. They conclude that the most likely model is bending due to the large-scale velocity of magnetised plasma in the ICM. NGC 6109 belongs to a small concentration of galaxies at one edge of the cluster Zw 1611+3717 \citep{ulrich78}, which itself lies in a supercluster extending over 60 Mpc. If large scale velocity structures from the ICM were responsible for the jet bends (and not accurately aligned with the jet direction), then the main jet would also be significantly bent once it reached beyond the galaxy's ISM (a few kpc), and this is not observed. 

A high fractional polarisation and well ordered magnetic field are often observed in bent jets and indicate shearing or compression of the radio source as it propagates through the intercluster medium. \cite{miley75} attribute the curved path of the radio trail in NGC 1265 to the orbit of the galaxy through the cluster and differential motions of the cluster medium. They report fractional polarisation as high as 60\% and well ordered magnetic field vectors along the radio emission. This is not the case for NGC 6109 since the fractional polarisation reaches 30\% at most and is usually around 10\%. 

The effect of galactic ram pressure and pressure gradients on jet deflection have been investigated (e.g. \citealp{jones79}; \citealp{begelman79}; \citealp{deyoung91}) in terms of the bending equation:

\begin{equation}
\frac{\rho_j v^2_j}{R_{\text{bend}}} = \frac{\rho_{\text{atm}} v^2_g}{R_p}
\end{equation}

\noindent where $\rho_j$ and $v_j$ are the density and velocity of the jet, $R_{\text{bend}}$ is the radius of curvature of the jet deflection, $\rho_{\text{atm}}$ is the density of the atmosphere and $v_g$ is the velocity of the parent galaxy. $R_p$ is the scale over which the ram pressure acting on the beam changes, and we set this equal to the width of the jet (i.e. $R_p \sim$ 1 kpc). Adopting $n_{\text{atm}}$ = 5 $\times 10^{-3}$ cm$^{-3}$ (see section 6.1), $v_g$ = 600 km s$^{-1}$ \citep{ulrich78}, $v_j$ = 1.5 $\times$ 10$^8$ m s$^{-1}$ and $R_{\text{bend}}$ = 3 kpc gives a jet density contrast $\rho_j / \rho_{\text{atm}} \approx$  2 $\times$ 10$^{-5}$. Therefore a hydrodynamical model is consistent with the tight bending if the jet is very light.

The rotation measure in the south of the doughnut component is suggestive of some external magnetic medium being responsible for, or contributing to the large scale bending seen in NGC 6109. Although the precession model can replicate the observed structure, it cannot explain complex RM structure only in the doughnut region. Instead we attribute the structure to an interaction between the radio plasma and some unknown magnetic density enhancement in the intergalactic medium. Evidence for the external magnetic gas is seen in the rotation measure and fractional polarisation maps, where the emission has been depolarised by the presence of structured Faraday rotating material. Figure~\ref{Fig:fpol1} shows alignment between the RM and the jet direction and a `cap' at the SE edge. It is possible that the counter-jet has lifted magnetised material and deposited it along its length and at the point where it turns, perhaps in a similar way to the cold gas uplifted in M 87 (\citealp{churazov01}). This could explain the high RM at the SE edge of the doughnut (the `cap'), however the RM along the main jet does not show a similar structure. If the main jet and the counter-jet both lift a similar amount of material the densities on the two sides will differ due to the Doppler factor of the lifted material, as RM$_{\text{jet}}$ = $\frac{1}{\delta}$ RM$_{\text{counter-jet}}$ (excluding RM due to the Galactic foreground). Our results however require a Doppler factor $\delta$ $\geq$ 7 to produce the observed RM structure, assuming RM$_{\text{galactic}}$ $\sim$ 20 rad m$^{-2}$. Such a high Doppler factor would be unexpected.

Under an interaction model, as the counter-jet turns at the collision site, compression of the jet leads to the formation of shocks, which can cause the brightening of the radio emission. South of the doughnut the magnetic field is circumferential, and appears to track the bending of the radio jet. The radio jet continues to deflect around a helical path, and high flux is seen to the NW of the structure, as a result of possible superposition of the jet with itself. The jet is then observed as diffuse emission to the NE. Here the magnetic field is radial as the jet spreads out into the external medium. Along the E edge of the diffuse emission, the magnetic field is longitudinal, possibly tracking the edge of the flow. The radial field lines observed through the centre of the doughnut could possibly be the result of compression due to the action of the external magnetic field. 

The external gas density of $n_e$ = 5 $\times$ 10$^{-3}$ cm$^{-3}$ at 10 arcsec from the core determined from the X-ray analysis can be used to provide an estimate for the magnetic field strength causing the high RM. The line of sight magnetic field B$_{\parallel}$ is calculated using the equation

\begin{equation}
RM = 812 \int n_e B_{\parallel} dl    
\end{equation}

\noindent where the rotation measure is measured in rad m$^{-2}$, $n_e$ is the electron density, in cm$^{-3}$, $B_{\parallel}$ is the line of sight magnetic field, in mG, and $dl$ is an element of the path length, in parsec. For RM = 200 rad m$^{-2}$, and $n_e$ = 5 $\times$ 10$^{-3}$ cm$^{-3}$ and assuming a path length of approximately 1 kpc, $B_{\parallel}$ = 50 $\mu$G, or 5 nT. This magnetic field strength is of the same order as that observed in the lobes of AGN and the value of $B_{eq}$ calculated in section 6. This lends support to the idea that the magnetic plasma could be remnants from a radio galaxy that passed close to NGC 6109, leaving behind a magnetised trail. This plasma would have to be well mixed with the IGM to produce RM. One difficulty this model faces is why an interaction with magnetised gas would cause the jet to continuously bend, rather than deflect and then continue on a linear path. A 180$^{\circ}$ jet reversal is observed in NGC 7385, but NGC 6109 requires a further push to complete the observed loop. A high residual vorticity in the trail could explain this; this trail would not be observed by the VLA, but could possibly be found by a low-frequency, high resolution map of the field by (e.g.) LOFAR. In addition, higher resolution (e.g. X-band VLA) observations would enable us to investigate substructure in the loop. If a gas-driven deflection is responsible, we would expect to see substructure transverse to the circumference of the doughnut, showing the effect of the environment on the flow (possibly with a large gradient in polarisation fraction from the inside to the outside of the doughnut). Far weaker radial effects would be expected if the flow is ballistic.

Given the difficulties with a ballistic model, it seems most likely that the unusual structure of NGC 6109 is due to an interaction. Further radio observations are required to investigate the nature of this interaction.

\section{Summary}

This study of NGC 6109 was undertaken in order to investigate how low redshift radio galaxies within the 3CRR sample interact with the external environment. VLA radio observations of this source reveal a remarkable `doughnut' component emitted on the opposite side of the core to a relatively straight jet. No infrared, optical or X-ray emission is associated with this unusual structure. Radio polarimetry shows a complex magnetic field structure across the doughnut, directed predominantly radial to the jet axis. In addition, In addition, high RM structures are found for this component. We summarise the conclusions of our investigation below:

\begin{enumerate}

\item The morphology of the doughnut might be explained ballistically provided the jet loses its integrity over the course of a single loop (precession period). This could be due to an external wind pushing against the component, which causes it to become diffuse and sweep back along the direction of the main jet but the absence of X-ray or optical emission suggests that the density of the external gas is very low. The bright feature at $\sim$7 kpc along the main jet is suggestive of an ejection event at the same time as the doughnut. For the system to lie fully within the galaxy atmosphere it must be at an angle $>$ 35$^{\circ}$ to the line of sight, but ballistic simulations find that the observed morphology requires much smaller line of sight angles. VLBI observations of the core may be able to detect proper motion associated with a binary black hole system causing such precession, but the only evidence for a ballistic model is the surprisingly circular morphology of the doughnut, as seen at 0.8$''$ resolution. \\

\item An interaction model where the path of the counter-jet is disrupted due to the external environment is possible. A simple hydrodynamical model would require a very light jet ($\rho_{\text{j}} / \rho_{\text{atm}} \sim 2 \times 10^{-5}$) to produce the tight doughnut structure.  The presence of strong RM structure to the south of the doughnut suggests an enhanced density or field in that region. Optical observations and H-$\alpha$ imaging would be useful to investigate whether there is any line-emitting gas coincident with the RM structure, which could lead to a deflection scenario similar to that proposed for NGC 7385 \citep{rawes15}. In addition, RM structure could be better discerned with lower frequency, high resolution observations (e.g. by LOFAR).\\

\end{enumerate}

\section{Acknowledgments}

JR acknowledges funding from STFC. The National Radio Astronomy Observatory is a facility of the National Science Foundation operated under cooperative agreement by Associated Universities, Inc. This research has made use of the NASA/IPAC Extragalactic Database (NED), which is operated by the Jet Propulsion Laboratory, California Institute of Technology, under contract with the National Aeronautics and Space Administration. The scientific results reported in this article include observations made with the \chandra\ X-ray Observatory.  We are grateful to the \chandra\ X-ray Center (CXC) for its support of \chandra\ and the {\sc ciao} software.  

\bibliographystyle{mnras}
\bibliography{refs}

\end{document}